\RequirePackage{fix-cm}
\documentclass{svjour3}                     
\smartqed  

\usepackage{graphicx}
\usepackage{bm}

\newcommand{\kn}{K$_{0.71}$Na$_{0.29}$Fe$_2$As$_2$}
\newcommand{\RuO}{RuOx }
\newcommand{\Ohm}{\Omega}
\newcommand{\dtime}[1]{\tau_{\mathrm{#1}} }
\newcommand{\AuEr}{\textbf{Au}:Er}
\newcommand{\tc}{$T_{\rm C}$}

\journalname{Journal of Low Temperature Physics}
\begin{document}


\title{Specific heat of K$_{0.71}$Na$_{0.29}$Fe$_2$As$_2$ at very low temperatures}


\titlerunning{Specific heat of K$_{0.71}$Na$_{0.29}$Fe$_2$As$_2$ at very low temperatures}        

\author{A. Reifenberger \and M. Hempel \and P. Vogt \and S. Aswartham \and M. Abdel-Hafiez \and V. Grinenko \and S. Wurmehl \and S.-L. Drechsler \and A. Fleischmann \and C. Enss \and R.
Klingeler}

\authorrunning{A. Reifenberger et al.} 

\institute{A. Reifenberger \and M. Hempel \and P. Vogt  \and A. Fleischmann \and C. Enss \and R. Klingeler \at
              Kirchhoff-Institut f\"ur Physik, Universit\"at Heidelberg, INF 227, D-69120 Heidelberg \\
              \email{andreas.reifenberger@kip.uni-heidelberg.de}           
           \and
           S. Aswartham \and M. Abdel-Hafiez \and V. Grinenko \and S. Wurmehl \and S.-L. Drechsler \at Leibniz Institute for Solid State Research, IFW Dresden, Helmholtzstr. 20, D-01069 Dresden}

\date{Received: date / Accepted: date}

\maketitle

\begin{abstract}
A commercially available calorimeter has been used to investigate the specific heat of a high-quality \kn\ single crystal. The addenda heat capacity of the calorimeter is determined in the temperature range $0.02 \, \mathrm{K} \leq T \leq 0.54 \, \mathrm{K}$. The data of the \kn\ crystal imply the presence of a large $T^2$ contribution to the specific heat which gives evidence of $d$-wave order parameter symmetry in the superconducting state. To improve the measurements, a novel design for a calorimeter with a paramagnetic temperature sensor is presented. It promises a temperature resolution of $\Delta T \approx 0.1 \, \mathrm{\mu K}$ and an addenda heat capacity less than $200 \, \mathrm{pJ/K}$ at $ T < 100 \,
\mathrm{mK}$.

\end{abstract}

\section{Introduction}
\label{intro}

Specific-heat studies down to very low temperatures provide insight into low-energy excitations of
the electronic, phononic, or magnetic subsystems of solids~\cite{Stewart}. Being directly linked to the entropy
changes, i.e. $\Delta S = \int (c_p/T) dT$, the specific heat $c_p$ is a valuable tool to study such excitations as well
as phase transitions. One example is the onset of superconductivity in BCS superconductors where a specific heat jump
$\Delta c_p = 1.43\gamma$\tc\ appears at \tc\ ($\gamma$ is the Sommerfeld coefficient). Well below \tc , the specific
heat gives direct access to the entropy of Cooper-pair breaking and in BCS superconductors it exponentially depends on the isotropic gap $\Delta$. To be more general, the specific heat measures the gap magnitude and structure and provides information on the pairing mechanism. Noteworthy, being a thermodynamic quantity the specific heat is sensitive to bulk properties which is in contrast to rather surface-sensitive methods such as ARPES or STM.

In this work, a commercially available device is applied for specific heat measurements of the unconventional superconductor \kn~\cite{Abdel13}. The data imply a large $T^2$-contribution to the specific heat well below \tc\ thereby evidencing
$d$-wave superconductivity in this material. The calibration of the calorimeter by measurements of high purity Ag however
indicates a strong Schottky-like increase of the addenda heat capacity which increases systematic errors below $\sim
60\, \mathrm{mK}$. In order to study materials with small heat capacity, i.e. with very small sample mass and/or
low specific heat, the design of a novel calorimeter is presented. The proposed calorimeter with paramagnetic temperature sensor and a SQUID-based readout is expected to have a much smaller addenda heat capacity of less than $200 \, \mathrm{pJ/K}$ for $ T < 100 \, \mathrm{mK}$
and promises a temperature resolution of $\Delta T \approx 0.1 \, \mathrm{\mu K}$.


\section{Experimental}
\label{experimental}

\kn\ single crystals were grown using KAs-flux described in Ref.~\cite{Abdel12}. A single crystal with mass $m=1.6 \, \mathrm{mg}$ was placed on a commercial calorimeter (heat capacity puck QD-P107H) from Quantum Design \cite{QD}. Apiezon N grease~\cite{Apiezon} (typically less than $0.5 \, \mathrm{mg}$) served as an adhesive between the sample and the measuring platform. This platform consists of a $3 \times 3 \times 0.25 \, \mathrm{mm^{3}}$ sapphire single crystal borne by Kapton strips. Two ruthenium oxide thick film resistors (\RuO resistors) attached to the sapphire platform serve as heater (with resistance $R_{\mathrm{Htr}}(T)$) and thermometer ($R_{\mathrm{Th}}(T)$).
Henceforth this setup is referred to as addenda. Both resistors are electrically contacted via Pt92W8-wires, which also define the thermal link between addenda and thermal bath. The calorimeter is mounted to the mixing chamber of a dilution refrigerator. For calibration purposes, the \RuO resistors' temperature dependencies $R_{\mathrm{Htr}}(T)$ and $R_{\mathrm{Th}}(T)$ were measured by a standard 4-wire sensing method using an AVS-47 AC resistance bridge. For the measurement of the heat capacity we applied a standard pulse-fitting method as described in literature \cite{Hwang}.

Heating power for the pulses is supplied using the analog voltage output of the data acquisition hardware NI-USB 6251 Box from National Instruments~\cite{NI} connected in series with a $4.7\, \mathrm{M\Ohm}$-resistance. A lock-in amplifier (Signal Recovery 7265 DSP) performing AC 4-wire sensing in the low $\mathrm{kHz}$-range is used to measure the temperature response with desired resolution both in time and amplitude. To suppress parasitic heating, the lock-in amplifier is galvanically detached from the platform and all wires attached to the platform are low-pass-filtered at room temperature with a cut-off frequency of $f_{3\mathrm{\,dB}} \approx 15\,\mathrm{MHz}$. The lock-in amplifier signal was calibrated against a carbon resistance thermometer placed at the mixing chamber; this thermometer had been calibrated against a fixed-point thermometer (SRD1000 from HDL~\cite{HDL}) and a noise thermometer (see Ref.~\cite{Netsch}).

\begin{figure*}\sidecaption
\resizebox{0.65\textwidth}{!}{\includegraphics*{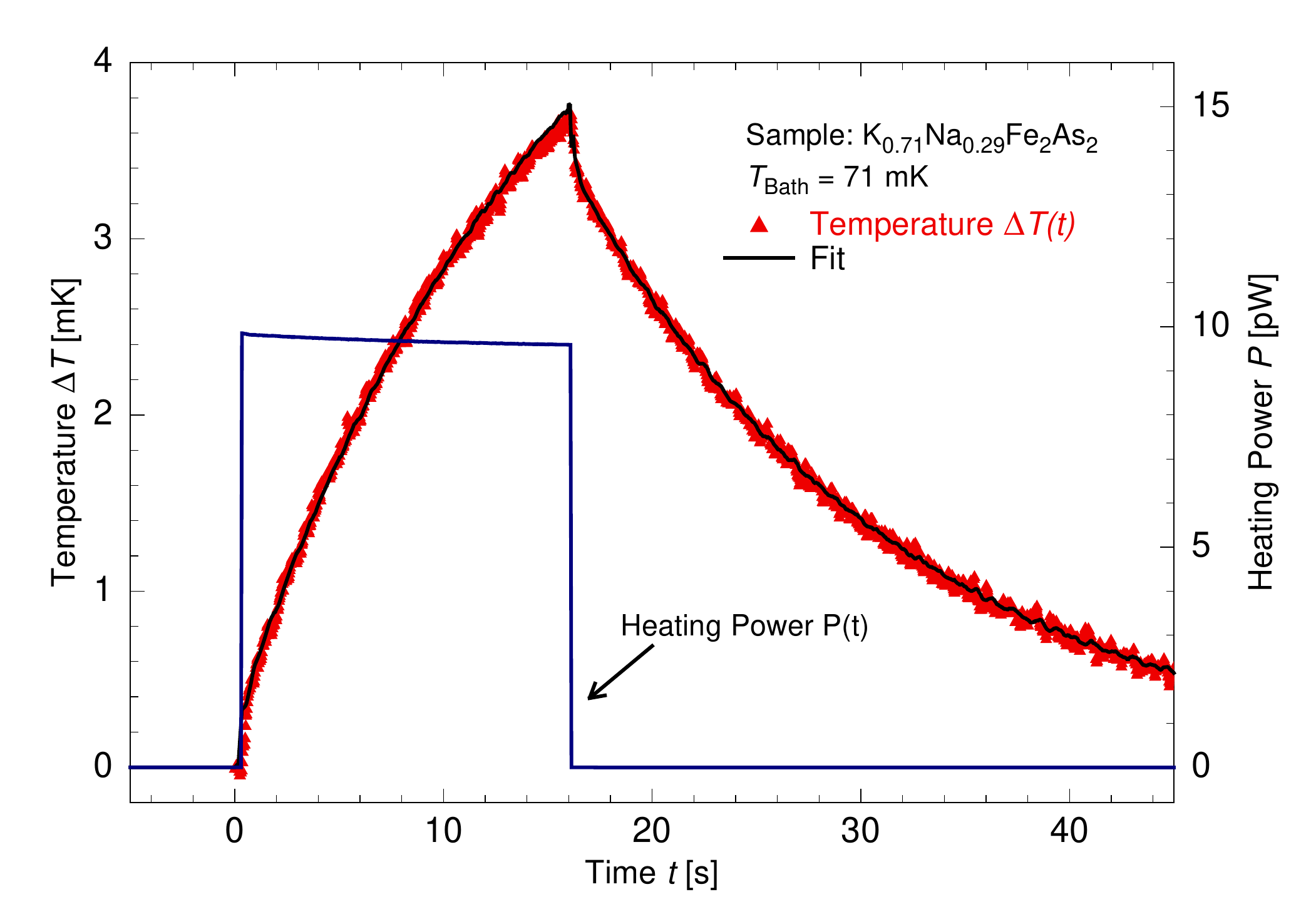}} \caption{A typical heat pulse and the associated temperature response measured with the set-up described in the text. A heat pulse starting at $t = 0 \, \mathrm{s}$ results in a temperature rise $\Delta T(t)$ followed by relaxation towards bath temperature. The data show two different relaxation processes with time constants $\dtime{1} \approx 15.7 \, \mathrm{s}$ and $\dtime{2} \approx 0.13 \, \mathrm{s}$ which arise from finite thermal couplings between sample and addenda and between addenda and thermal bath.}
\label{SinglePulse}
\end{figure*}

A typical heat pulse and the corresponding temperature response is shown in Fig.~\ref{SinglePulse}. After the heat pulse, the temperature dependence $\Delta T(t)$ shows two different relaxation processes associated with the thermal links between addenda and sample and between addenda and thermal bath, respectively. The temperature relaxation is described by two exponentials with relaxation times $\dtime{1}$ and $\dtime{2}$. The black curve represents the model fit applied to describe the pulse $\Delta T(t)$ from which fitting parameters are extracted to determine the total heat capacity of sample and addenda.

\begin{figure*}\sidecaption
\resizebox{0.65\textwidth}{!}{\includegraphics*{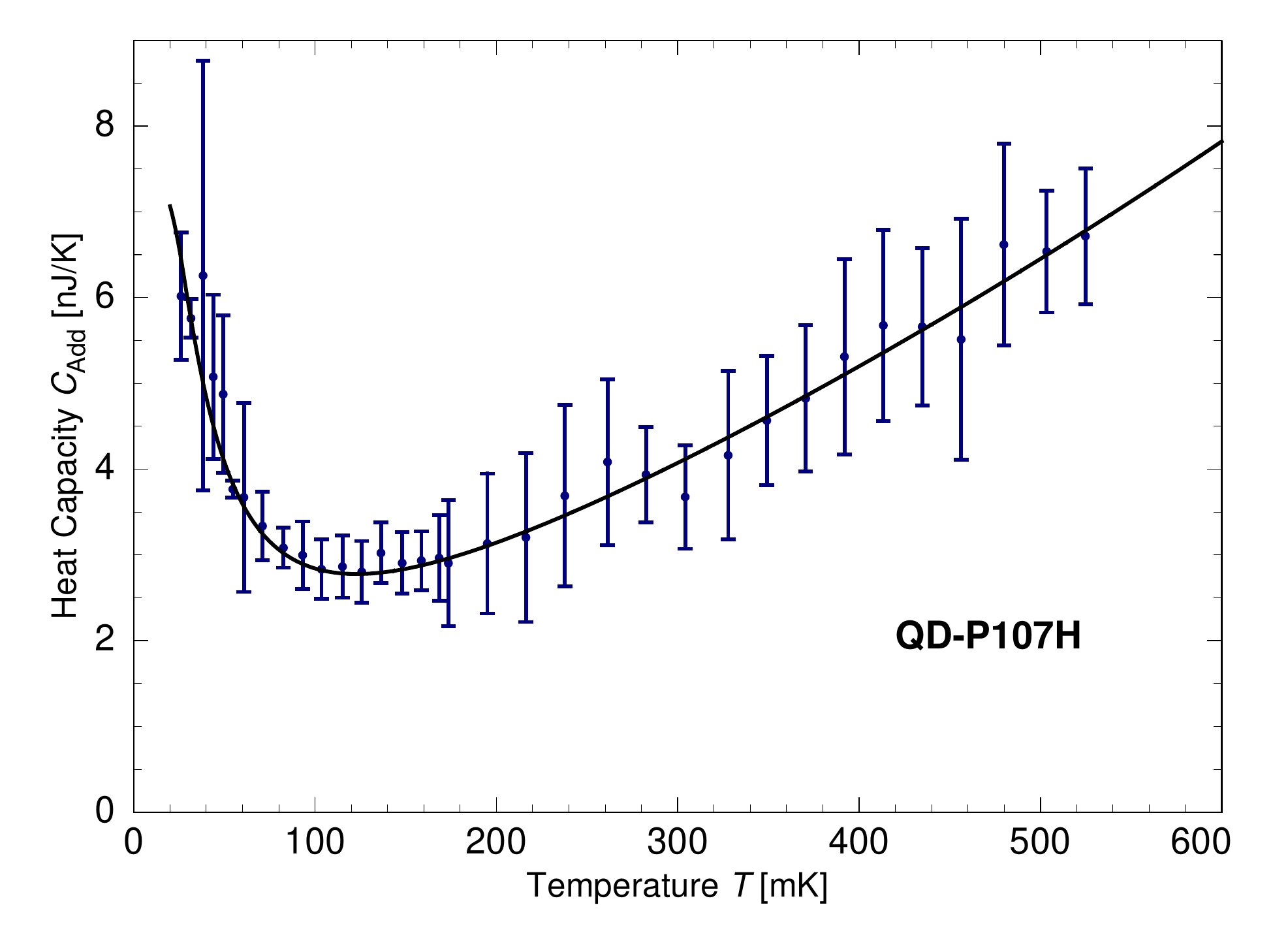}} \caption{Addenda heat capacity of QD-P107H with
statistical error bars. The black line represents a fit to the data (see in the text).}
\label{Addenda}
\end{figure*}

For calibration, a silver sample with nominal purity of $99.9999\,\%$ and mass $m = 9.95 \pm 0.05 \, \mathrm{mg}$ is used. Magnetic susceptibility measurements in a commercial SQUID magnetometer (QuantumDesign MPMS-XL5~\cite{QD}) revealed no detectable magnetic impurities in the temperature range $2 \, \mathrm{K} < T < 300 \,\mathrm{K}$ that might contribute to the specific heat of this
sample. From theoretical calculations one derives the electronic specific heat of silver~\cite{Ziman}. The Debye coefficient is derived from the Debye temperature from Ref.~\cite{Smith}. The silver specific heat is considered to be \begin{equation}c_{\mathrm{Ag}} = \gamma_{\mathrm{el}} \, T + \beta \, T^3 \;\end{equation} with $\gamma_{\mathrm{el}} = 644 \,\mathrm{\mu J mol^{-1} K^{-2}}$  and $\beta = 167 \,\mathrm{\mu J mol^{-1} K^{-4}}$~\cite{Ziman,Smith}. An additional correction due to the Apiezon N grease specific heat is taken into account as well~\cite{Schink}. The resulting addenda heat capacity obtained after subtracting the silver heat capacity is shown in Fig.~\ref{Addenda}. The error bars shown indicate the statistical error of typically ten individual pulses measured.

A continuous $C_{\mathrm{Add}}(T)$-curve is obtained  by approximating the experimental data by means of an appropriate arbitrary empirical function $C_{\mathrm{Add}}(T) = a+b T + c T^3 + d \tanh \left(e / (T-f)\right)$, with arbitrary parameters $a = 0.012 \, \mathrm{nJ/K}$, $b = 11.5 \, \mathrm{nJ/K^2}$, $c= 3 \, \mathrm{nJ/K^4}$, $d = 7\, \mathrm{nJ/K}$, $e = 0.022\, \mathrm{K}$ and $f = 0.0099\, \mathrm{K}$. The result of this procedure is shown by the black line in Fig.~\ref{Addenda}. The data show a nearly linearly increasing addenda heat capacity above a minimum at around $T \approx 120 \, \mathrm{mK}$ and a Schottky-like increase below this minimum. While no further details of the thermometer composition have been
communicated to the authors, one may however assume that Ru nuclear moments with $I = 5/2$ present in the thermometer or the Pt92W8-wires cause the observed Schottky-like behaviour~\cite{Volokitin,Ho}.


\section{Measurements on \kn}

\begin{figure*}\sidecaption
\resizebox{0.65\textwidth}{!}{\includegraphics*{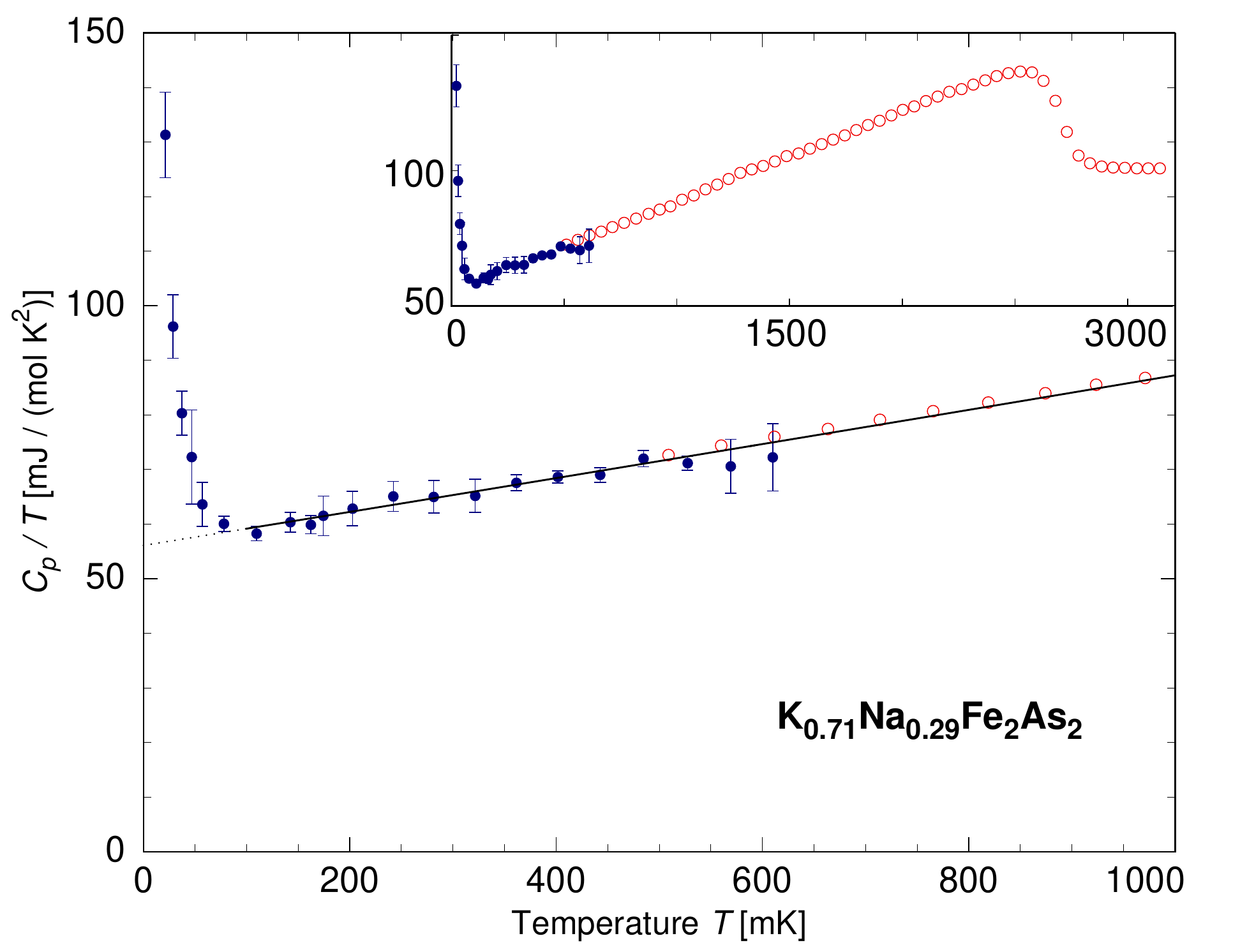}} \caption{Low-temperature specific heat $c_p/T$ of
\kn\ vs. temperature at 20\,mK $\leq T \leq 1$\,K. Data have been obtained using the low-temperature calorimeter described in the text (filled circles) and by means of a Quantum Design PPMS (open circles). The inset shows an extended temperature range up to $T>$ \tc . The lines reflect a fit to the data according to Eq.~(\ref{gleichung}).}
\label{KNa122}       
\end{figure*}

Although $s^{\pm}$-pairing has been suggested for the entire class of Fe-based superconductors (e.g., Ref.~\cite{Mazin}, and references therein), the nature of superconductivity in these materials is still under debate and specific heat studies are one of the major experimental tools to address this issue~\cite{Ding,Welp,Stockert,Abdel12a,Hardy}. Here, single-crystalline \kn\ was studied by means of the device described above in the temperature regime between $0.02 \, \mathrm{K}$ and $0.65 \, \mathrm{K}$ (Fig.~\ref{KNa122}). The presence of superconductivity in the crystal was confirmed by measurements of the volume ac susceptibilities ($\chi^{\prime}$ and $\chi^{\prime \prime}$) which yields the superconducting transition temperature $T_{\mathrm{c}} = 2.75 \, \mathrm{K}$, and $4\pi \chi^{\prime} (T = 2 \, \mathrm{K}) = -1$~\cite{Abdel13}. In the inset of Fig.~\ref{KNa122}, where the specific heat at higher temperatures up to $T \approx 3 \, \mathrm{K}$ obtained by a Quantum Design PPMS system is shown, the associated specific heat jump $\Delta c_p$ at \tc\ is clearly visible~\cite{Abdel13}. The specific heat jump at \tc\ amounts to  $\Delta c_p/ T_{\rm C} \approx 40 \, \mathrm{mJ/mol K^2}$.  At low temperatures the data show a linear-in-$T$ decrease of $c_p/T$, i.e. $c_p \propto T^2$, which is superimposed by a Schottky-like contribution below $\sim 100$\,mK. Note, that at $20 \, \mathrm{mK}$ the sample heat capacity is larger than the addenda contribution by a factor of two (see Fig.~\ref{Addenda}). This ratio strongly increases upon heating to, e.g., 10 at $100 \, \mathrm{mK}$.

The experimentally observed  $T^2$-dependence of the specific heat well below \tc\ evidences quasi-particle excitations
near line node(s), i.e. the superconducting gap in \kn\ is zero at least at one $k$-point of one Fermi surface sheet.
The data hence imply nodal superconductivity, i.e. either $d$- or $s_{\pm}$-wave symmetry of the pairing state. For
quantitative analysis of the low-temperature specific heat, the specific heat is fitted by
\begin{equation}
c_{p} = \gamma_r T + \alpha T^2 + \beta_3 T^3 \, \label{gleichung}
\end{equation}
which yields the residual Sommerfeld-coefficient $\gamma_{\mathrm{r}} = 55.2 \, \mathrm{mJ mol^{-1} K^{-2}}$, the
quasi-particle contribution $\alpha = 35.4 \, \mathrm{mJ mol^{-1} K^{-3}}$ and the lattice contribution $\beta = 0.556
\, \mathrm{mJ mol^{-1} K^{-4}}$. Note, that the lattice contribution was defined from the normal state behaviour (for
details see the supplement of Ref.~\cite{Abdel13}). The large $T^2$-contribution arises mainly from quasi-particle excitations near line node(s) in \kn . A quantitative estimate in Ref.~\cite{Abdel13} suggests $d$-wave superconductivity in \kn .

\section{Design of a novel magnetic thermometry-based calorimeter}

\begin{figure*}[b!]\sidecaption
\resizebox{0.65\textwidth}{!}{\includegraphics*{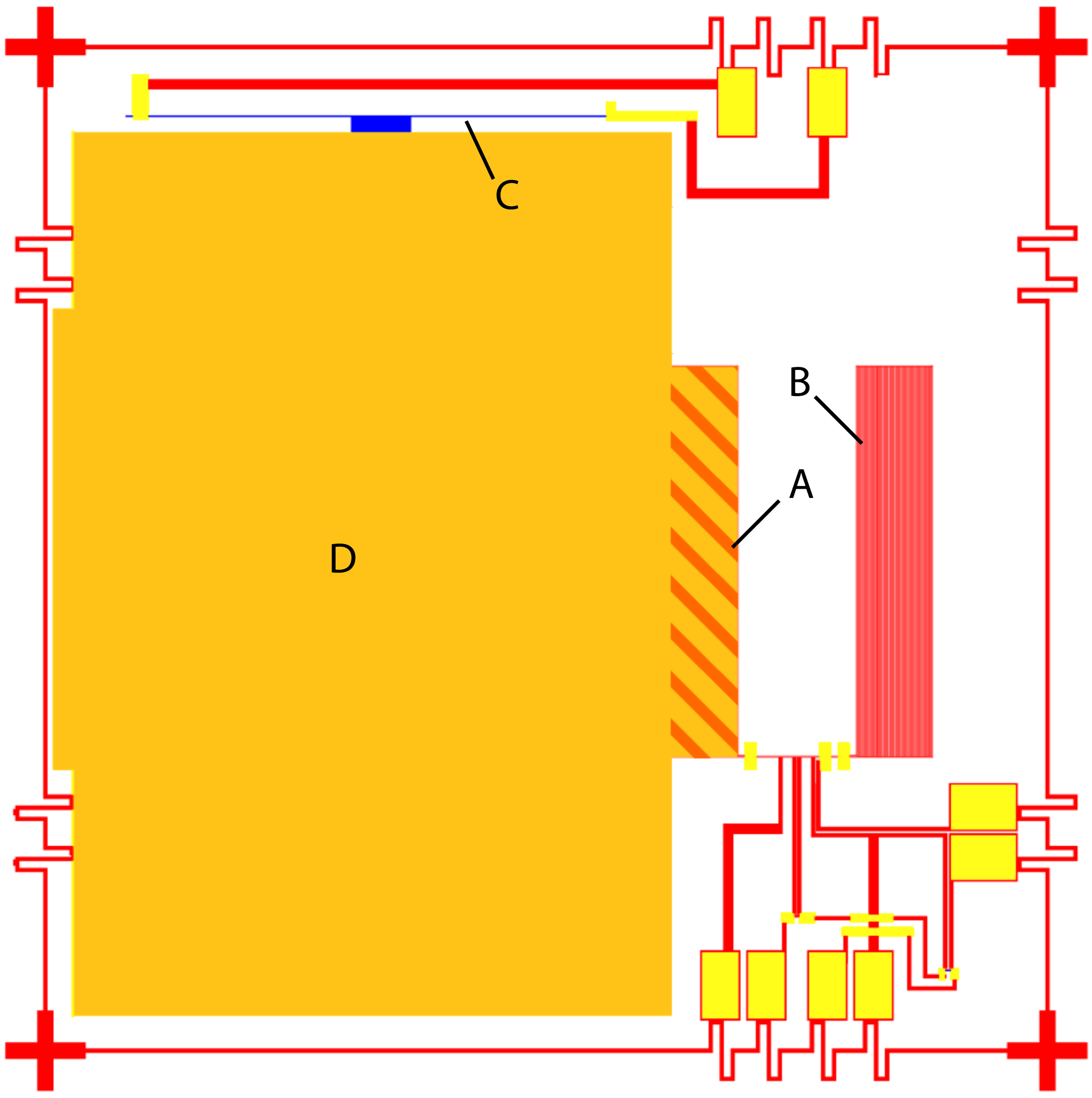}} \caption{Design of a calorimeter with paramagnetic temperature sensor and SQUID-based readout for measurements at ultra-low temperatures. The device is based on a sapphire chip with two
superconducting coils (A, B), a AuPd resistor (C), a \AuEr ~ temperature sensor (A) and a gold layer for placing the sample on (D). Note, that coil A is hidden by the overlay.} \label{HeatCapChip}
\end{figure*}

\begin{figure*}\sidecaption
\resizebox{0.65\textwidth}{!}{\includegraphics*{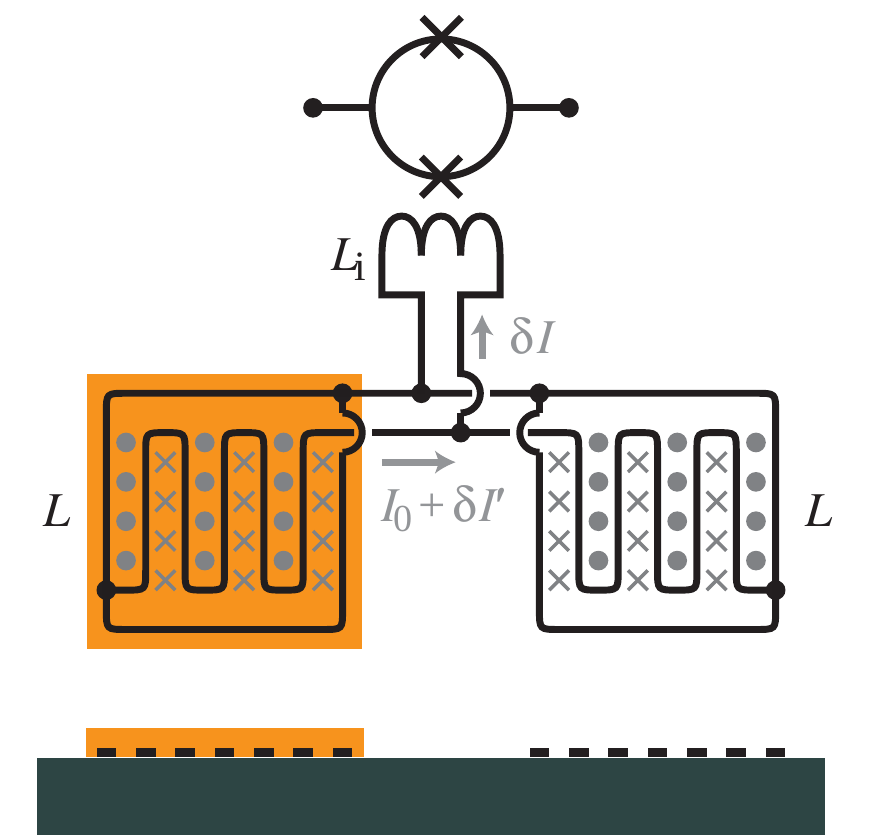}} \caption{Schematic sketch of the magnetic
thermometer. Two meander-shaped superconducting coils with inductance $L$ are placed on the chip (see
Fig~\ref{HeatCapChip}). A persistent current $I_{\mathrm{0}}$ produces an inhomogeneous magnetic field (dots and crosses). The sensor material (orange), separated by a thin insulating layer, is placed above one of the
pick-up coils. A change of its magnetisation leads to a screening current $\delta I^{\prime}$ in the meander-shaped coils and $\delta I$ in the input inductance $L_{i}$ of the SQUID, respectively (see text for details). From Ref.~\cite{Fleischmann2005}.} \label{Meander_Scheme}
\end{figure*}

The device discussed so far has limitations at temperatures below $50 \, \mathrm{mK}$. Firstly, the observed
Schottky-contribution to the addenda heat capacity yields a lower limit to the resolution of the obtained specific
heat. Secondly, in order to avoid self-heating only low voltages can be applied to the \RuO thermometer
resulting in a nonsatisfying temperature resolution. A third problem arises when the $\dtime{2}$ effect \cite{Hwang,Schwall} becomes
increasingly dominant at low temperatures. The model applied for data analysis assumes a temperature-independent
thermal link $K_\mathrm{1}$ during each heat pulse. However, this precondition is not met anymore when, due to a large
$\dtime{2}$ effect, the sample platform temperature difference between the heating and cooling process is too large. To
address these issues, a calorimeter with paramagnetic temperature sensor and SQUID-based readout is suggested as shown in Figs.~\ref{HeatCapChip} and
\ref{Meander_Scheme}.

Paramagnetic metallic sensors are well established low-temperature thermometers for micro-calorimeters (see Ref.~\cite{Enss2000,Fleischmann2005}). These devices have been developed and utilized for high resolution particle detection. Basic elements of magnetic calorimetry-based detectors are absorbers for the particles and paramagnetic sensors made of Er-doped Au (\AuEr) whose magnetisation obeys a Curie-like behaviour. In a small magnetic field, the magnetisation strongly depends on temperature. Temperature changes due to absorption of particles can be detected by a highly sensitive SQUID-based read-out. The details of the underlying physics of the sensor material and the detection scheme have been investigated thoroughly in the last decade~\cite{Enss2000,Fleischmann2005,Fleischmann2009,Pies}.


The new calorimeter design depicted in Fig.~\ref{HeatCapChip} features a \AuEr-sensor with an internal signal rise time as fast as $100\, \mathrm{ns}$ at $30\,\mathrm{mK}$ \cite{Fleischmann2009} as a central element. The samples will be placed on a $1\, \mathrm{\mu m}$ thin gold layer (sample platform) with a usable area of $3 \times 4.5 \, \mathrm{mm^{2}}$ micro-fabricated on a sapphire substrate ($5 \times 5 \times 0.3 \, \mathrm{mm^{3}}$). A AuPd film resistor (C in Fig.~\ref{HeatCapChip}) placed on the sapphire is used to generate a heat pulse. The heater is in electronic contact with the sample platform (D) to allow for fast heat flow. The gold layer may also be used to attach samples via ultrasonic bonding rather than Apiezon N grease. This will strongly decrease the undesired $\dtime{2}$ effect. Thermal equilibration is achieved via Au bonds between the sample platform and the thermal bath. By varying the number of Au bonds, we can adjust the thermal link $K_\mathrm{1}$ depending on the expected heat capacity of the sample under investigation.

To detect the temperature response of a heat pulse, the sample platform is electronically coupled to the sensor material (A) of the thermometer. This sensor material is placed in a small magnetic field produced by a persistent current in meandering superconducting coils (B and underneath A) with inductance $L = 40 \, \mathrm{nH}$. Together with the input coil of a dc SQUID, these coils form a superconducting flux transformer. The temperature read-out scheme is depicted in Fig.~\ref{Meander_Scheme}. The SQUID itself is placed next to the calorimeter chip and the coils are connected via superconducting Niobium bonds. A gradiometric read out scheme of the paramagnetic temperature sensor is employed in order to minimize signal fluctuations caused by external magnetic disturbances.


Based on experience with cryogenic particle detectors~\cite{Fleischmann2005,Pies}, thermodynamic
properties of the presented setup including the specific heat and the magnetisation of the sensor material can be
predicted reliably using numerical methods. Assuming typical values for the persistent current ($100\, \mathrm{mA}$)
and the doping level of Er ($300 \, \mathrm{ppm}$), the contribution of the sensor material \AuEr ~to the addenda heat
capacity is estimated to be $30 - 110 \, \mathrm{pJ/K}$ in the temperatures range $100 - 10 \, \mathrm{mK}$. Furthermore, the
sample platform has an electronic specific heat of $C_\mathrm{Au} = 1\, \mathrm{nJ/K^{2}} \, T$~\cite{Martin1973}.
Other specific heat contributions can be neglected due to the small amount of material used (AuPd) or their small
intrinsic heat capacities (Nb, phonons in the sapphire substrate). In total, we expect a 40-fold smaller addenda heat capacity compared to the
discussed commercial device. The temperature resolution of the new calorimeter can be predicted taking into account
geometric factors (sensor volume $V_\mathrm{s}$, meander-pitch $p$), the circuitry of the meanders and SQUID
performance parameters. With the presented design we expect a temperature resolution of $\Delta T \approx 0.1 \mathrm{\mu K}$ with an integration time of several~$\mathrm{m s}$, which is far better and faster than the \RuO resistance thermometer of the presently used commercial device.

\section{Summary}

A commercially available calorimeter was calibrated and used for specific heat studies of \kn\ down to 20\,mK in a dilution refrigerator. A large $T^2$-term in the specific heat implies nodal superconductivity whose nature is found to be consistent with $d$-wave by quantitative analysis in Ref.~\cite{Abdel13}. At lowest temperatures, the commercial device is of
limited use only and the design for a calorimeter based on magnetic thermometry is presented. The micro-fabricated magnetic calorimeter promises temperature resolution of $\Delta T \approx 0.1 \, \mathrm{\mu K}$ and addenda heat capacity less than $200 \, \mathrm{pJ/K}$ for $ T < 100 \, \mathrm{mK}$.

\begin{acknowledgements}
Valuable discussions with S. Kempf, C. Pies, and A. Reiser are gratefully acknowledged. This work was supported by the DFG through projects KL1824/6, WU595/3-1, BU887/15-1, En299/5-1 and by the European Community Research Infrastructures under the FP7 Capacities Specific Program, MICROKELVIN project number 228464.

\end{acknowledgements}



\end{document}